\documentstyle[twocolumn,floats,aps]{revtex}
\newcommand{\be}{\begin{equation}}
\newcommand{\ee}{\end{equation}}
\newcommand{\bea}{\begin{eqnarray}}
\newcommand{\eea}{\end{eqnarray}}

\begin{document}
\draft 
\twocolumn[\hsize\textwidth%
\columnwidth\hsize\csname@twocolumnfalse\endcsname
\title{
{Phase Space Reduction and the Instanton Crossover in (1+1)-Dimensional Turbulence}}
\author{ L. Moriconi$^{a)}$ and G.S. Dias$^{b)}$}
\address{$a)$ Instituto de F\'\i sica, Universidade Federal do Rio de Janeiro \\C.P. 68528, Rio de Janeiro, RJ --- 21945-970, Brasil}
\address{$b)$ Centro Brasileiro de Pesquisas F\'\i sicas, Rua Xavier Sigaud, 150\\ Rio de Janeiro, RJ --- 22290-180, Brasil}
\maketitle
\begin{abstract}  
We study (1+1)-dimensional turbulence in the framework of the Martin-Siggia-Rose field theory formalism. The analysis is focused on the asymptotic behaviour at the right tail of the probability distribution function (pdf) of velocity differences, where shock waves do not contribute. A BRS-preserving scheme of phase space reduction, based on the smoothness of the relevant velocity fields, leads to an effective theory for a few degrees of freedom. The sum over fluctuations around the instanton solution is written as the expectation value of a functional of the time-dependent physical fields, which evolve according to a set of Langevin equations. A natural regularization of the fluctuation determinant is provided from the fact that the instanton dominates the action for a finite time interval. The transition from the turbulent to the instanton dominated regime is related to logarithmic corrections to the saddle-point action, manifested on their turn as multiplicative power law corrections to the velocity differences pdf.
\end{abstract}
\vskip1pc] 
\narrowtext

\section{Introduction}
As commonly acknowledged through the vast amount of numerical investigations of turbulence, intermittency -- in short, the deviation of pdf's tails from simple gaussian behaviour -- turns out to be a phenomenon intimately tied to the number of spatial dimensions. The basic physical picture behind this observation is that long lived coherent structures, which play roughly the role of local reservoirs of conserved quantities, like energy, interact more intensely in lower dimensions, perturbing in a relevant way the cascade process between large and small length scales. It is in this sense mostly that Burgers model of compressible turbulence in (1+1) dimensions \cite{burgers} has been a fruitful ground for theoretical investigation. Intermittency is associated there with the existence of shock waves, which are described in the inviscid limit, $\nu \rightarrow 0$, by large positive velocity gradients $\partial_x u \sim u_0^2/ \nu$ with support in vanishingly small regions of sizes $\sim  \nu / u_0$. Kolmogorov's theory of the inertial range \cite{kolmogorov}, which gives structure functions $S_q(r) \equiv \langle |u(r)-u(0)|^{q} \rangle \sim r^{2q/3}$, has long been known not to match the scaling behaviour found in the Burgers model, $S_q(r) \sim r$ \cite{burgers,saffman}, obtained straightforwardly from the fact that the turbulent one-dimensional fluid may be depicted as a dilute gas of shock waves, interpolated by smooth velocity fields
\cite{saffman,yakhot}.

A considerable boost in the understanding of one-dimensional turbulence has been achieved in the last few years 
\cite{yakhot,gotoh,polyakov,gura,bouchad,balko,weinan,bec-frisch}, with the help of a number of completely different approaches, among them non-perturbative field-theory methods. Most efforts have been concentrated on the determination of the velocity differences or velocity derivatives pdfs. It is now clear that the right and left tail of the asymmetric pdfs deserve separate treatments, with the left tail results being the subject of some controversy. 

Our aim in this work is to define an effective theory for the computation of the right tail pdf, retaining the most important degrees of freedom in the analysis. Initial attempts on this problem were done by Polyakov \cite{polyakov} and Gurarie and Migdal \cite{gura}, who have found leading order expressions. Our approach closely follows the instanton method employed by the latter. The computation of perturbations around the instanton will be carried out here, where, after a natural regularization procedure of the fluctuation determinant, subleading logarithm corrections for the saddle-point action will emerge. The effective theory will be defined in such a way as to preserve the fundamental BRS symmetry of the Martin-Siggia-Rose field theory \cite{msr,zinn}, related to general properties like renormalization and the existence of a statistical stationary state (proved to hold in the class of ``purely dissipative Langevin equations"). 

This paper is organized as follows. In sec. II we introduce the basic equations to be studied and the Martin-Siggia-Rose path-integral formulation. The BRS-preserving scheme of phase space reduction is performed, and the instanton configuration is defined. Fluctuations around the saddle-point fields are then considered in order to compute the subleading logarithm corrections. In sec. III we show that, up to first order, viscosity does not modify the previous results, in accordance with a well-defined inviscid limit. In sec. IV we comment on our findings and discuss possible steps of further research.

\section{Phase space reduction in the Martin-Siggia-Rose formalism}
Let us imagine that the one-dimensional fluid is kept in a state of sustained turbulence, by means of the action of external random forces.
Our starting point is therefore the stochastic version of Burgers model, defined by the Navier-Stokes equation
\be
\partial_t u + u \partial_x u = \nu \partial^2_x u + f  \ . \  \label{psr1}
\ee
The force $f = f(x,t)$ is taken to be a random gaussian variable, with zero mean and two-point correlation function
\be
\langle f(x,t) f(x',t') \rangle = D(x-x') \delta(t -t')  \ , \  \label{psr2}
\ee
where
\be
D(x-x')= D_0 \exp [- (x-x')^2 /L^2 ]  \ . \  \label{psr3}
\ee
Above, $L$ sets the macroscopic length scale where energy pumping is assumed to occur. As $L \rightarrow \infty$ and the viscosity vanishes, an intermediate range of wavenumbers forms -- the inertial range -- where scaling behaviour comes into play. Arbitrary $N$-point correlation functions of the velocity field may be computed in principle through the Martin-Siggia-Rose functional \cite{msr,zinn},
\be
Z= \int D \hat u D u D \bar c D c \exp( i S ) \ , \ \label{psr4}
\ee
with
\bea
S&&=\int dx dt \hat u( \partial_t u + u \partial_x u - \nu \partial^2_x u) \nonumber \\
&&+ { i \over 2} \int dx dx' dt \hat u(x,t) \hat u(x',t) D(x-x')
\nonumber \\
&&+\int dx dt \bar c ( \partial_t c + c \partial_x u + u \partial_x c
-\nu \partial^2_x c) \ . \ \label{psr5}
\eea
Above, $c(x,t)$ and $\bar c(x,t)$ are Grassmann variables. These fields enter into the mathematical apparatus as a way to account for the non-trivial jacobian defined in the continuum time formulation \cite{zinn,cardy}. The Martin-Siggia-Rose field theory approach may be used to define a perturbative expansion involving powers of the convection term, where the jacobian's role is just to cancel tadpole diagrams. However, it is important to note that the perturbative procedure is essentially the Wyld expansion \cite{wyld}, plagued with the well-known insurmountable infrared divergencies as $L \rightarrow \infty$ \cite{orzsag}.

We are interested to study the statistics of the velocity differences $\zeta \equiv u(\rho/2,0) - u(-\rho/2,0)$ in a non-perturbative fashion. The pdf for this observable quantity may be computed as \cite{polyakov,gura}
\be
P(\zeta) = {1 \over {2 \pi}}
\int_{- \infty}^{\infty} d \lambda \exp(i \lambda  \zeta) Z( \lambda) \ , \ \label{psr6}
\ee
where
\bea
Z(\lambda)&&=\langle \exp \{ - i \lambda [ u(\rho/2,0) - u(-\rho/2,0)] \}
\rangle \nonumber \\
&&= \int D \hat u D u D \bar c D c \exp( i S_\lambda ) \label{psr7}
\eea
and
\be
S_\lambda = S -  \lambda [ u(\rho/2,0) - u(-\rho/2,0)] \ . \ \label{psr8}
\ee
We will consider the analytic mapping $\lambda \rightarrow i \lambda$ in the next computations. Our results, to be obtained in the large $\lambda$ limit, will be, therefore, concerned with the right tail of the pdf, where shock waves do not contribute. The precise mathematical statement is that after the analytic mapping the limit $\lambda \rightarrow \infty$ is related to $\zeta \rightarrow \infty$ in eq. (\ref{psr7}), whenever the right tail pdf decays faster than $\exp(-c \zeta)$ for any $c>0$. To prove it, we write (\ref{psr7}) as $Z(i \lambda) \propto \int_{-\infty}^{\infty} d\zeta P(\zeta) \exp(\lambda \zeta)$. When $\lambda \rightarrow \infty$, we will have
\be
Z(i \lambda) \sim \exp[ \ln P( \zeta_0)+\lambda \zeta_0] \ . \ \label{comm1}
\ee
Above, we used the usual saddle-point approximation to get the resulting asymptotic expression. The parameter $\zeta_0$ is obtained as the solution of the saddle-point equation,
\be
{d \over {d \zeta}} P(\zeta) = - P(\zeta) \lambda \ . \ \label{comm2}
\ee
Observe that from the knowledge of the asymptotic behaviour of
$Z(i \lambda)$ we are allowed to obtain $P(\zeta_0)$. All we need, thus, is that $\zeta_0 \rightarrow \infty$ when $\lambda \rightarrow \infty$, in order to get the form of the right tail pdf.

Taking, for instance, $P(\zeta) \sim \exp( - c \zeta^\alpha)$ at the right tail, it is clear that $Z(i \lambda)$ converges only for $\alpha >1$. Furthermore, we find $\zeta_0 \sim \lambda^{ 1 / (\alpha -1)}$, which grows with $\lambda$, so that the large $\lambda$ limit is related in fact to the form of the right tail pdf. This is the state of affairs in Burgers turbulence, where, as we will see, $\alpha = 3$.

The action $S$ is BRS-invariant, according to the following variations:
\be
\delta \hat u = 0 \ , \
\delta u = \epsilon c \ , \
\delta c = 0 \ , \
\delta \bar c = - \epsilon \hat u \ . \ \label{psr9}
\ee
Above, $\epsilon$ is a constant Grassmann parameter. It is crucial, in any kind of computational strategy, to work in a BRS symmetry preserving scheme, which ensures the interpretation of the path integral $Z(\lambda)$ in terms of some stochastic dynamical system. This is the fundamental guiding principle to be followed, in order to get a reduced form of the Martin-Siggia-Rose functional. Let us define the power series
\bea
&&u(x,t) = \sum_{p=0}^\infty \sigma_p (t) x^p \ , \
\hat \sigma_p (t) = \int dx x^p \hat u(x,t) \nonumber \\
&&c(x,t) = \sum_{p=0}^\infty \eta_p (t) x^p \ , \
\bar \eta_p(t) = \int dx x^p \bar c(x,t) \ . \ \label{psr10} 
\eea
Applying (\ref{psr9}) to the above relations, we immediately get the ``reduced-BRS" transformations
\be
\delta \hat \sigma_p = 0 \ , \
\delta \sigma_p = \epsilon \eta_p \ , \
\delta \eta_p = 0 \ , \
\delta \bar \eta_p = - \epsilon \hat \sigma_p \ . \ \label{psr11}
\ee
The implicit physical motivation in (\ref{psr10}) is that the relevant velocity fields involved in the computation of the pdf's right tail are smooth. In other words, the above transformation to a discrete set of dynamical variables is closely related to the Taylor expansion of the stochastic force term in the Burgers equation, which is a meaningful procedure 
in the absence of shock waves.

A phase-space reduction may be implemented by truncating the series expansions (\ref{psr10}) at some arbitrary order. Furthermore, if one wants to study the explicit role of viscous terms, it is necessary to truncate the series beyond first order. The simplest choice is to take the expansion of $u(x,t)$ up to the second order, which gives
\bea
&& S_\lambda =  \int dt \{  \hat \sigma_0 ( \dot \sigma_0 + \sigma_0 \sigma_1
-2 \nu \sigma_2)+ \hat \sigma_1 ( \dot \sigma_1 + \sigma_1^2
+2 \sigma_0 \sigma_2) \nonumber \\
&& + \hat \sigma_2 ( \dot \sigma_2 + 3 \sigma_1 \sigma_2)
+i {D_0 \over 2} (\hat \sigma_0^2 - {2 \over L^2} \hat \sigma_0 \hat \sigma_2
+ {2 \over L^2} \hat \sigma_1^2
+ {3 \over  L^4} \hat \sigma_2^2) \nonumber \\
&& + \bar \eta_0(\dot \eta_0 +\eta_0 \sigma_1-2 \nu \eta_2 + \eta_1 \sigma_0)
+ \bar \eta_1(\dot \eta_1 +2 \eta_1 \sigma_1 +2 \eta_2 \sigma_0 \nonumber \\
&&+ 2 \eta_0 \sigma_2) +\bar \eta_2(\dot \eta_2 +3 \eta_1 \sigma_2+3\eta_2 \sigma_1)
-i \delta(t) \lambda \rho \sigma_1 \} \ . \ \nonumber \\
\label{psr12}
\eea
For $\lambda =0$, the above action is invariant under the reduced-BRS transformations, as it should be. The saddle-point equations, meaningful in the large $\lambda$ limit, are
\bea
&& \dot \sigma_0 + \sigma_0 \sigma_1 -2 \nu \sigma_2 + i D_0 \hat \sigma_0 -i {D_0 \over  L^2} \hat \sigma_2 = 0  
\ , \ \nonumber \\
&& \dot \sigma_1 + \sigma_1^2 + 2 \sigma_0 \sigma_2 + i {{2D_0} \over L^2} \hat \sigma_1 = 0 \ , \ \nonumber \\
&& \dot \sigma_2 + 3 \sigma_1 \sigma_2 - i {D_0 \over L^2} \hat \sigma_0
+i {{3 D_0} \over L^4} \hat \sigma_2 = 0 \ , \ \nonumber \\
&& \dot {\hat \sigma_0} - \hat \sigma_0 \sigma_1-2 \hat \sigma_1 \sigma_2-\bar \eta_0 \eta_1 -2 \bar \eta_1 \eta_2 = 0 \ , \ \nonumber \\
&& \dot {\hat \sigma_1} - \hat \sigma_0 \sigma_0 - 2 \hat \sigma_1 \sigma_1
-3 \hat \sigma_2 \sigma_2 - \bar \eta_0 \eta_0 -2 \bar \eta_1 \eta_1 \nonumber \\
&& -3 \bar \eta_2 \eta_2 + i \lambda \rho \delta (t) = 0 \ , \  \nonumber \\
&& \dot {\hat \sigma_2} + 2 \nu \hat \sigma_0-2 \hat \sigma_1 \sigma_0 -3 \hat \sigma_2 \sigma_1 - 2 \bar \eta_1 \eta_0 -3 \bar \eta_2 \eta_1 =0 \ , \ \nonumber \\
&& \dot \eta_0 + \eta_0 \sigma_1 - 2 \nu \eta_2 + \eta_1 \sigma_0=0
\ , \ \nonumber \\
&&  \dot \eta_1 + 2  \eta_1 \sigma_1 + 2 \eta_2 \sigma_0 +
 2 \eta_0 \sigma_2 = 0 \ , \ \nonumber \\
&& \dot \eta_2 + 3 \eta_1 \sigma_2   + 3 \eta_2 \sigma_1 = 0 
\ , \ \nonumber \\
&& \dot {\bar \eta_0} - \bar \eta_0 \sigma_1 - 2 \bar \eta_1 \sigma_2 = 0  \ , \ \nonumber  \\
&& \dot {\bar \eta_1} - \bar \eta_0 \sigma_0 -2 \bar \eta_1 \sigma_1
-3 \bar \eta_2 \sigma_2 = 0 \ , \ \nonumber \\
&& \dot {\bar \eta_2} +2 \nu \bar \eta_0- 2 \bar \eta_1 \sigma_0
-3 \bar \eta_2 \sigma_1= 0 \ . \ \label{psr13}
\eea
The exact saddle-point equation for $\hat u (x,t)$ shows it evolves with ``negative viscosity". Thus, it is  necessary, in order to solve (\ref{psr13}), to impose the boundary condition $\hat \sigma_p (0^+) =0$. The instanton found by Gurarie and Migdal \cite{gura} corresponds to take all variables but $\hat \sigma_1$ and $\sigma_1$ equal to zero (note that corrections appear at larger orders), which yields
\bea
&&\sigma_1^{(0)}(t)={ d \over {dt}} \ln R(t) \ , \ \nonumber \\
&&\hat \sigma_1^{(0)}(t)=i \lambda \rho R(t)^2 \ , \ \label{psr14}
\eea
where
\be
R(t)= (1 - t / \tau)^{-1} \ , \
\tau =\left ( { L^2 \over {D_0 \lambda \rho}} \right )^{1 \over 2}
\ . \ \label{psr15}
\ee
The fields in the Martin-Siggia-Rose action may be defined now as perturbations departing from the saddle-point solutions, through the replacement
\bea
&&\sigma_1 \rightarrow  \sigma_1^{(0)} + \sigma_1 \ , \ \nonumber \\
&&\hat \sigma_1 \rightarrow  \hat \sigma_1^{(0)} + \hat \sigma_1 \ . \ \label{psr16}
\eea
Substituting these expressions in (\ref{psr12}), we obtain a factorized form for $Z(i \lambda)$, expressed as the product of the dominant saddle-point contribution and a correction term, related to fluctuations around the instanton:
\be
Z( i \lambda) = \exp[i S_{0}(\lambda)] Z_1(\lambda) \ , \ \label{psr17}
\ee
where
\bea
&&S_{0}(\lambda) = -i {{2 D_0^{1/2}}\over {3 L}} (\lambda \rho)^{3/2} \ , \ \nonumber \\
&&Z_1(\lambda) = \int \prod_{p=0}^2 D \hat \sigma_p D \sigma_p
D \bar \eta_p D \eta_p \exp[i S_1(\lambda)] \label{psr18}
\eea
and
\bea
S_1(\lambda) &&= \int_{- \infty}^0 dt \{  \hat \sigma_0 ( \dot \sigma_0 + \sigma_0 \sigma_1 + \sigma_0 \sigma_1^{(0)}
-2 \nu \sigma_2) \nonumber \\
&&+ \hat \sigma_1 ( \dot \sigma_1 + \sigma_1^2
+2 \sigma_1^{(0)} \sigma_1
+2 \sigma_0 \sigma_2) +\hat \sigma_2 ( \dot \sigma_2
+ 3 \sigma_1 \sigma_2 \nonumber \\
&&+ 3 \sigma_1^{(0)}
\sigma_2 )+i {D_0 \over 2} (\hat \sigma_0^2 - {2 \over L^2} \hat \sigma_0 \hat \sigma_2 + {2 \over L^2} \hat \sigma_1^2 + {3 \over L^4} \hat \sigma_2^2)  \nonumber \\
&&+ \bar \eta_0(\dot \eta_0 +\eta_0 \sigma_1
+\eta_0 \sigma_1^{(0)}-2 \nu \eta_2 + \eta_1
\sigma_0) \nonumber \\
&&+ \bar \eta_1(\dot \eta_1 +2 \eta_1 \sigma_1+
2 \eta_1 \sigma_1^{(0)}+2 \eta_2 \sigma_0 + 2 \eta_0
\sigma_2) \nonumber \\
&&+\bar \eta_2(\dot \eta_2 +3 \eta_1 \sigma_2+3\eta_2 \sigma_1
+3 \eta_2 \sigma_1^{(0)}) \nonumber \\
&&+\hat \sigma_1^{(0)} ( \sigma_1^2 + 2 \sigma_0
\sigma_2)\} \ . \ \label{psr19}
\eea
Observe, as usual, that there are no first order terms in the definition of $S_1(\lambda)$, 
since they vanish due to the saddle-point equations. The path integral $Z_1 (\lambda)$ has a simple and interesting interpretation. We may write
\be
Z_1 (\lambda) = \langle \exp[ i \int_{- \infty}^0 dt \hat \sigma_1^{(0)} ( \sigma_1^2 + 2 \sigma_0\sigma_2)] \rangle \ , \ \label{psr20}
\ee
where the above average is determined by the following stochastic (Langevin) equations,
\bea
&&\dot \sigma_0 + \sigma_0 \sigma_1^{(0)}+\sigma_0 \sigma_1 -2 \nu \sigma_2 = f_0(t) \ , \ \nonumber \\
&&\dot \sigma_1+2 \sigma_1 \sigma_1^{(0)}+ \sigma_1^2 +2 \sigma_0 \sigma_2 = f_1(t) \ , \ \nonumber \\
&&\dot \sigma_2+3 \sigma_2 \sigma_1^{(0)} + 3 \sigma_1 \sigma_2 = f_2 (t)
\ . \ \label{psr21}
\eea
The time-dependent random forces are correlated as $\langle f_i (t) f_j (t') \rangle = \delta(t-t') \Omega_{ij}$,
with
\be
\Omega = {D_0 \over L^2}
\left[
\matrix{ L^2 & 0 & -1  \cr
         0 & 2  & 0\cr
         -1  & 0 & 3L^{-2}  } \right] \ . \  \label{psr22} \qquad
\ee
The mapping between eqs. (\ref{psr18}) and (\ref{psr20}) is exact. The proof follows from an explicit construction: starting with the stochastic system given by eqs. (\ref{psr21}) and (\ref{psr22}), we find, through the Martin-Siggia-Rose formalism, that the expectation value in eq. (\ref{psr20}) may be precisely written as eq. (\ref{psr18}). This exact mapping is a central point in the analysis.

Taking into account only the linear terms in the Langevin equations (\ref{psr21}), in accordance with the theory of quadratic fluctuations around the saddle-point, and considering the limit of vanishing viscosity, we are able to get the exact solution, with initial condition $\sigma_p (-T)=0$ (where it is meant $T \rightarrow  \infty$)
\bea
\sigma_p (t) &&= \exp[-(p+1)\int_0^t dt_1 \sigma_1^{(0)} (t_1)] \nonumber \\
&& \times \int_{-T}^t dt_2 f_p(t_2)  \exp[(p+1)\int_0^{t_2} dt_3 \sigma_1^{(0)} (t_3)] \nonumber \\
&& =R(t)^{-(p+1)} \int_{-T}^t dt' f_p(t') R(t')^{(p+1)} \ . \  \label{psr23}
\eea
A direct application of Wick's theorem shows that internal contractions of $\sigma_1^2 + 2 \sigma_0 \sigma_2$ vanish. Only contractions among fields defined at different times will produce non-vanishing results. We have, thus, expanding the exponential in (\ref{psr20}), the normal-ordered expression
\bea
Z_1( \lambda)&&= \sum_{n=0}^\infty {i^n \over {n!}} 
\int_{-T}^0 dt_p \hat \sigma_1^{(0)}(t_p) \nonumber \\
&& \times \langle \prod_{p=1}^n:[\sigma_1^2(t_p) + 2 \sigma_0(t_p) \sigma_2(t_p)]: \rangle
\ . \ \label{psr24}
\eea
There are divergences in the computation of $Z_1(\lambda)$ as
$T/ \tau \rightarrow  \infty$. Taking the expansion up to second order, it is not difficult to get
\be
Z_1 (\lambda) \sim \ln(T / \tau) \ . \ \label{psr25}
\ee
Reasoning in terms of the cumulant expansion, we find that the above logarithmic behaviour clearly implies that $Z_1$ scales like a power of $T/ \tau$. The regularization of this divergent result is a subtle point. Note that for times $t \ll - \tau$, we expect fluctuations to be governed by the full Martin-Siggia-Rose action describing Burgers turbulence, since the instanton becomes in this time range essentially irrelevant. Thus, it is natural to suppose that time splits into two regions. We may define in fact a time instant $t_0$ so that for $t < t_0$ (region I) fluctuations of $\sigma_p$ are described by pure Burgers turbulence, where as the interval $t_0 < t < 0$ (region II) is dominated by the instanton, where it is consistent to apply the theory of quadratic fluctuations. The velocity field in region II is assumed to be smooth, that is
\be
\sigma_0 \gg \sigma_1 \rho \gg \sigma_2 \rho^2 \gg ... \ . \ 
\label{psr26}
\ee
As we will see, this condition is satisfied by the solutions of the Langevin equations, provided that $L \gg \rho$. To find $t_0$, just take any of the Langevin equations, as the one for $\sigma_1(t)$, for instance. Looking at eq. (\ref{psr21}), we are then interested to know at what time the quadratic term $\sigma_1^2$ becomes more relevant than $\sigma_1 \sigma_1^{(0)}$. A pragmatical way to compare the role of these two terms is just to replace $\sigma_1$ by $\sqrt{\langle  \sigma_1^2 \rangle}$. That is, we have to compute
\be
g(t) \equiv { {\sigma_1^{(0)}} \over \sqrt{ \langle \sigma_1^2 \rangle} }
= { {R(t)} \over { \tau \sqrt{ \langle \sigma_1^2 \rangle}} } \ . \ \label{psr27}
\ee
The time instant $t_0$ is defined by the condition $g(t_0) \sim 1$. Using (\ref{psr23}) we get, in the limit $T / \tau \rightarrow \infty$,
\be
\langle  \sigma_1^2 \rangle = { {2 \tau D_0} \over {3 L^2}} R(t)^{-1} \ . \ \label{psr28}
\ee
Assuming now $|t_0/\tau| \gg 1$, which holds in the asymptotic limit of large $\lambda$, we find
\be
t_0 \sim -({L^2 \over D_0})^{{1 \over 3}} \ . \ \label{psr29}
\ee
The basic conclusion drawn from the above arguments is that we are enforced to solve (\ref{psr21}) for an arbitrary set of initial conditions at $t=t_0$. Let in this way $\bar \sigma_p \equiv \sigma_p(t_0)$ represent the initial value of the physical fields in region II. In order to match regions I and II, we write
\bea
Z_1( \lambda) &&= \int_C
d \bar \sigma_0 d \bar \sigma_1 d \bar \sigma_2 P( \bar \sigma_0, \bar \sigma_1, \bar \sigma_2) \nonumber \\
&& \times \langle \exp[ i \int_{t_0}^0
dt \hat \sigma_1^{(0)} ( \sigma_1^2 + 2 \sigma_0 \sigma_2) ] \rangle
\ . \ \label{psr30}
\eea
The subindex $C$ denotes integration over smooth configurations of the velocity field, and $P(\bar \sigma_0, \bar \sigma_1, \bar \sigma_2)$ is the pdf for the set of dynamical variables at time $t_0$, which encodes all the information about fluctuations developed in region I. According to the picture of Burgers turbulence as a dilute gas of kinks of the velocity field \cite{yakhot}, the imposition of smoothness in the integration is attained when the coordinates $x= \pm \rho/2$ defined in the evaluation of velocity differences are placed in the region between shocks. Therefore, a physical interpretation here is to regard $|t_0|$ as the mean time spent at an arbitrary point until a shock wave crosses it.

The solutions of the Langevin eqs. (\ref{psr21}), which contain now the explicit dependence on initial conditions, are given by
\be
\sigma_p (t) = \xi_p(t) + \left [ {{R (t_0)} \over {R (t)}} \right ]^{p+1} 
\bar \sigma_p \ , \ \label{psr31}
\ee
with
\be
\xi_p(t)=R(t)^{-(p+1)} \int_{t_0}^t dt' f_p(t') R(t')^{(p+1)} \ . \ \label{psr32}
\ee
It is important to check, as discussed before, the smoothness condition (\ref{psr26}) for the above solutions. We obtain
\bea
\langle \sigma_p (t)^2 \rangle &&= \langle \xi_p (t)^2 \rangle
+  \left [ {{R(t_0)} \over {R(t)}} \right ]^{2(p+1)} 
\langle \bar \sigma_p^2 \rangle \nonumber \\
&&= \Omega_{pp} R(t)^{-2(p+1)} \int_{t_0}^t dt' R(t')^{2(p+1)} \nonumber \\
&&+ \left [ {{R(t_0)} \over {R(t)}} \right ]^{2(p+1)}
\langle \bar \sigma_p^2 \rangle \ , \ \label{psr33}
\eea
where $\langle \bar \sigma_p^2 \rangle$ is the average over initial conditions. From pure dimensional analysis, we have 
\be
\langle \bar \sigma_p^2 \rangle \sim (L^{(1-p)}/t_0)^2 \ . \ \label{psr34}
\ee
Now, since $ dR(t) /dt > 0$, we get, for the integral in (\ref{psr33})
\bea
\int_{t_0}^t dt' R(t')^{2(p+1)} 
&&= \int_{t_0}^t dt' R(t')^{2(p+2)} R(t')^{-2} \nonumber \\
&&> R(t)^{-2} \int_{t_0}^t dt' R(t')^{2(p+2)} \ . \ \label{psr35}
\eea
Applying this inequality back to (\ref{psr33}) and taking into account that
\be
{\Omega_{pp} \over \Omega_{p+1,p+1} } \sim
{{\langle \bar \sigma_p^2 \rangle } \over 
{\langle \bar \sigma_{p+1}^2 \rangle}} \sim L^2 \gg \rho^2 \label{psr36}
\ee
and also
\be
\left [ {{R(t_0)} \over {R(t)}} \right ]^{2(p+2)} <
\left [ {{R(t_0)} \over {R(t)}} \right ]^{2(p+1)} \ , \ \label{psr37}
\ee
we get
\be
\langle \sigma_p(t)^2 \rangle \gg 
\langle  \sigma_{p+1}(t)^2 \rangle \rho^2 \ , \ \label{psr38}
\ee
which is the smoothness condition (\ref{psr26}) stated in a more precise way.

The fluctuating field $\xi(t)$ is a linear functional of the force field, with correlation function 
\bea
&&A_{ij}(t_1,t_2) \equiv \langle \xi_i (t_1)\xi_j(t_2) \rangle =
\nonumber \\
&&= \tau R(t_1)^{-(i+1)}R(t_2)^{-(j+1)} { {\Omega_{ij}} \over {(i+j+1)}}
[ F_{i+j}(t_2) \Theta(t_1-t_2) \nonumber \\
&& + F_{i+j}(t_1) \Theta(t_2-t_1)] \ , \ \label{psr39}
\eea
where
\bea
F_{p}(t)&&=(p+1) \tau^{-1} \int_{t_0}^t dt' R(t')^{(p+2)} \nonumber \\
&&= R(t)^{(p+1)}-|\tau / t_0|^{(p+1)} \ . \ \label{psr40}
\eea
Note that $A_{ij}(t_1,t_2)$ is a positive definite operator. The expression for $Z_1 (\lambda)$ becomes
the product of two factors:
\be
Z_1( \lambda) = Z_{1a} (\lambda) Z_{1b}(\lambda) \ , \ \label{psr41}
\ee
where
\bea
&&Z_{1a} (\lambda) = \langle \exp[ i \int_{t_0}^0
dt \hat \sigma_1^{(0)} ( \xi_1^2 + 2 \xi_0 \xi_2) ] \rangle \nonumber \\
&&= \langle \exp[ - \int_{t_0}^0
dt \int_{t_0}^0 dt' \xi_i(t) B_{ij}(t,t') \xi_j(t') ] \rangle \ , \  \nonumber \\
&&Z_{1b}(\lambda)= \int_C
d \bar \sigma_0 d \bar \sigma_1 d \bar \sigma_2 P( \bar \sigma_0, \bar \sigma_1, \bar \sigma_2) \nonumber \\
&& \times \exp[ \int_{t_0}^0 dt \int_{t_0}^0 dt' J_i(t) M_{ij}(t,t') J_k (t') \nonumber \\
&&-{{t_0^2} \over 3} (\bar \sigma_1^2 + 2 \bar \sigma_0 \bar \sigma_2 )]
\ , \ \nonumber \\ \label{psr42}
\eea
with
\bea
&& B(t,t') = \lambda \rho R(t)^2 \delta(t-t')
\left[
\matrix{ 0 & 0 & 1  \cr
         0 & 1  & 0\cr
         1  & 0 & 0  } \right]  \ , \ \qquad  \nonumber \\
&& J = 2 \lambda \rho
\left[
\matrix{ R(t)^{-1} R(t_0)^3 \bar \sigma_2 \cr
         R(t_0)^2 \bar \sigma_1 \cr
         R(t)R(t_0) \bar \sigma_0 } \right]  \ , \ \label {psr43} \qquad
\eea
and
\be
M_{ij}(t,t')=
[(A^{-1}+4B)^{-1}]_{ij}(t,t') \ . \ \label{psr44}
\ee

We establish now some arguments in order to show that $Z_{1b}$ has a finite limit as $\tau \rightarrow 0$. Two cases have to be distinguished here, depending on $B$ being or not a positive definite operator. In the second order truncation we have been discussing, $B$ is not positive definite. In this situation, we were able only to devise a perturbative analysis.
We write
\be
M = {1 \over {A^{-1}+4 B}}=[1-4AB+(4AB)^2+...]A \ . \ \label{psr45}
\ee
Considering the first order approximation to the above series, a computer-assisted evaluation (it is necessary to examine a large number of terms) shows that both
\be
\int_{t_0}^0 dt_1 \int_{t_0}^0 dt_2 J_i(t_1)A_{ij}(t_1,t_2)J_j(t_2) \label{psr46}
\ee
and
\bea
&&\int_{t_0}^0 dt_1 \int_{t_0}^0 dt_2 \int_{t_0}^0 dt_3 \int_{t_0}^0 dt_4 \nonumber \\
&& \times J_i(t_1)A_{ij}(t_1,t_2)B_{jk}(t_2,t_3)A_{kl}(t_3,t_4)J_l(t_4) \label{psr47}
\eea
are finite when $\tau \rightarrow 0$. We conjecture that the same behaviour holds at any order in perturbation theory.
On the other hand, if $B$ is a positive definite operator, as it happens for the truncation of (\ref{psr10}) at first order, it is possible to address a proof of the above conjecture. At first, we observe that due to the fact that $A$ and $B$ are both real, positive definite and symmetric operators, inner products satisfy to
\be
\langle  \Psi ,( A^{-1}+4B)^{-1} \Psi  \rangle  < \langle  \Psi ,( A+{1 \over 4} B^{-1}) \Psi  \rangle \ . \ \label{psr48}
\ee
To understand why this is so, just define $\Phi = ( A^{-1}+4B)^{-1} \Psi $. Therefore,
\bea
&&\langle  \Psi ,( A+{1 \over 4} B^{-1}) \Psi  \rangle  =  \langle \Phi,( A^{-1}+4B)( A+{1 \over 4}B^{-1})\nonumber \\
&& \times ( A^{-1}+4B) \Phi \rangle
= 3\langle \Phi,( A^{-1}+4B) \Phi \rangle \nonumber \\
&&+ {1 \over 4} \langle \Phi, A^{-1} B^{-1} A^{-1} \Phi \rangle + 16 \langle \Phi, B A B \Phi \rangle \nonumber \\
&&=3\langle  \Psi ,( A^{-1}+4B)^{-1}  \Psi  \rangle
+ {1 \over 4} \langle A^{-1} \Phi ,B^{-1} A^{-1} \Phi \rangle \nonumber \\
&&+ 16 \langle B \Phi , A B \Phi \rangle > \langle  \Psi ,( A^{-1}+4B)^{-1}  \Psi  \rangle \ . \ \label{psr49}
\eea
This theorem may be applied now to the $JMJ$ term contained in the definition of $Z_{1b}$:
\bea
&& \int_{t_0}^0 dt \int_{t_0}^0 dt' J_i(t) M_{ij}(t,t') J_j (t') \nonumber \\
&&< \int_{t_0}^0 dt \int_{t_0}^0 dt' J_i(t) [A +{1 \over 4} B^{-1}]_{ij}(t,t') J_j (t') \ . \ \label{psr50}
\eea
We get in fact a finite limit for the rhs of the above expression, in the limit $\tau \rightarrow 0$.

It is interesting to discuss a possible physical interpretation for the result that $Z_{1b}( \lambda)$ does not depend on $\lambda$ in the asymptotic limit. Note that: i) the function $Z_{1b}(\lambda)$ is actually defined in terms of the coupling between fluctuations around the instanton solution and the initial conditions for the $\sigma_i$'s at time $t_0$. If we had fixed $\sigma_i(t_0)=0$, for instance, then $Z_{1b}( \lambda)$ would not depend on $\lambda$, as it can be easily derived from eqs. (\ref{psr42}) and (\ref{psr43}); ii) recalling that $t_0$ is the mean time spent until a shock wave crosses an arbitrary point in the one-dimensional space, we may regard $|t_0|$ in an equivalent way as the ``memory time" associated to large and positive fluctuations of $\zeta$ (the ``right tail" region). In other words, considering $P(\zeta_2|\zeta_1;t_2-t_1)$ -- the conditional pdf to observe $\zeta_2$ at time $t_2$ if $\zeta_1$ was observed at time $t_1$ -- we have that for large values of $\zeta_2$ and for $t \geq |t_0|$ it holds 
\be
P(\zeta_2|\zeta_1; t) \simeq P(\zeta_2) \ , \ \label{comm3}
\ee
that is, the conditional pdf becomes the pdf itself at the right tail, for $t \geq |t_0|$. Now, as the large $\zeta$ limit is related to the large $\lambda$ limit, we find, from i) and ii), that the basic physical reason for $Z_{1b}(\lambda)$ not to depend on $\lambda$ (in the asymptotic limit) is the absence of memory effects from fluctuations which took place in times
far enough in the past.

We have, thus, to concentrate our attention on $Z_{1a}$. Since only quadratic terms are involved in the definition of $Z_{1a}$, we find
\be
Z_{1a} = \exp[ -{1 \over 2} {\hbox{Tr}} \ln ( 1+4 A B ) ] \ . \ \label{psr51}
\ee
Expanding the logarithm as
$\ln ( 1+4 A B ) = 4AB -8(ABAB)+ ...$,
and taking into account that ${\hbox{Tr}} (AB) = 0$, the first non-zero 
contribution gives, in the limit $\tau \rightarrow 0$, and after a
lenghty computation,
\be
Z_{1a} = \exp[ 4{\hbox{Tr}}(ABAB) ] \sim \exp[ c \ln(|t_0 / \tau|) ] \ , \ \label{psr53}
\ee
with $c=224/45 \simeq 5$. Since $\tau \sim \lambda^{-1/2}$,
we get
\be
P( \zeta) \sim \zeta^c \exp( - \zeta^ {3} ) \ . \label{psr54}
\ee
This is our central result. A careful analysis of the instanton crossover has led ultimately to the power law correction
shown in eq. (\ref{psr54}), for the asymptotic form of the velocity differences pdf at the right tail. 

\section{Viscosity Effects}
In order to study viscosity effects, all we have to do, as it follows from eq. (\ref{psr21}), is to perform the replacement $f_0 \rightarrow f_0 +2 \nu
\sigma_2$, that is,
\be
f_0(t) \rightarrow f_0(t) + 2 \nu R(t)^{-3} \int_{t_0}^t dt' f_2(t') R(t')^3+
2 \nu \bar \sigma_2 \ . \ \label{ve1}
\ee
We have now,
\bea
&&\langle f_0(t) f_0(t') \rangle  = \nonumber \\ 
&&=\Omega_{00} \delta (t-t')
+2 \nu \Omega_{02} R(t')^3 R(t)^{-3} + {\cal{O}}(\nu^2) \nonumber \\
&&\langle f_0(t) f_2(t') \rangle = \nonumber \\
&&= \Omega_{02} \delta (t-t')
+2 \nu \Omega_{22} R(t')^3 R(t)^{-3}\Theta(t-t') \ . \
\label{ve2}
\eea
The above definitions may be used to compute corrections to the correlation functions $A_{ij}(t,t')$. We get, up to first order,
\be
{\hbox{Tr}} \ln ( 1+4 A B ) = 4 {\hbox{Tr}}(AB)\sim { {\nu t_0} \over L^2} \ . \ \label{ve3}
\ee
We find, thus, that viscosity does not lead to logarithm corrections, and that the limit $\nu \rightarrow 0$ is a well-defined one, at least to first non-trivial order in perturbation theory.

\section{Conclusions}
We studied the instanton crossover problem in the field-theoretical approach to Burgers turbulence. With $t_0 \equiv
-(L^2 / D_0)^{1/3}$, this crossover is characterized by the transition from fully developed turbulence, at times 
$t < t_0$, to the instanton dominated regime, in the interval $ t_0 < t < 0$, where quadratic 
fluctuations may be integrated out. The essential outcome of our analysis is that pdf's tails get multiplicative power law corrections to the leading, saddle-point expressions. Such phenomenon, which is hard to detect numerically, is likely to occur in other intermittent systems. 

We note that some questions are left open, which require even more labourious computations. As already observed, the operator $B$, introduced in eq. (\ref{psr43}), may be positive definite or not, according to the order where the truncation of (\ref{psr10}) is done. Since the sign of the parameter $c$ (defined in eq. (\ref{psr53})) is positive if $B$ is positive definite, but the converse is not necessarily true, it is clear that as higher order truncations are considered, we expect that i) $B$ becomes positive definite or not, with $c \rightarrow \bar c \geq 0$, or ii) $B$ becomes non-positive, with $c \rightarrow \bar c < 0$. We do not have enough information yet to decide which one is the correct choice; further work is necessary on this matter. Also, it is worth mentioning that power law corrections may be model-dependent, as strongly suggested by the leading order saddle-point results themselves, related to the specific form of the force-force correlation function.

A natural question is if analogous results hold for the left tail pdf, where the Martin-Siggia-Rose formalism may as well be employed \cite{balko}. The techniques discussed here are adequate to study this problem, a subject deserved for future investigations. Even though the path integral approach is direct and appealing, it relies on heavy perturbative computations performed on truncated series. It would be interesting, and perhaps useful, to know how the power law corrections could be derived from some alternative analysis of randomly forced Burgers turbulence (after completion of this work we became aware  of ref. \cite{tribe}, where power law corrections are rigourously obtained in the case of decaying Burgers turbulence).

\acknowledgements
This work was partially supported by CNPq.

\end{document}